\documentclass[12pt]{article}
\usepackage{a4wide}
\usepackage[dvips]{graphicx}
\usepackage{epsf}                        
\usepackage{amssymb}
\usepackage{latexsym}                    
\usepackage{amsfonts}                    

\makeatletter
\@addtoreset{equation}{section}
\makeatother

\begin{document}

\rightline{FPAUO-11/03}
\rightline{CERN-PH-TH/2011-183}

\vspace{1.5truecm}

\centerline{\LARGE \bf Dielectric 5-Branes and Giant Gravitons in ABJM}
\vspace{1.3truecm}

\centerline{
      {\large \bf Mario Herrero${}^{a}$},
      {\large \bf Yolanda Lozano${}^{a,b,}$}\footnote{E-mail address:
                                  {\tt ylozano@uniovi.es}},
    {\large \bf Marco Picos${}^{a,}$}\footnote{E-mail address:
                                  {\tt picosmarcos@uniovi.es}}                              
    }
                                                            
\vspace{.4cm}

\centerline{{\it ${}^a$ Department of Physics,  University of Oviedo,}}
\centerline{{\it Avda.~Calvo Sotelo 18, 33007 Oviedo, Spain}}

\vspace{.4cm}
\centerline{{\it ${}^b$Theory Group, Physics Division, CERN,}}
\centerline{{\it CH-1211 Geneva 23, Switzerland}}

\vspace{1truecm}

\centerline{\bf ABSTRACT}
\vspace{.5truecm}

\noindent
We construct a supersymmetric NS5-brane wrapped on a twisted 5-sphere expanding in the $CP^3$ in $AdS_4\times CP^3$,  with D0-brane charge. This configuration provides a realization of the stringy exclusion principle in terms of giant D0-branes. In the maximal case the twisted 5-sphere reduces to a $CP^2$ and its energy can be accounted for both by a bound state of $k$ D4-branes wrapping the $CP^2$ and a bound state of $N$ D0-branes, a realization on the gravity side of the symmetry of Young diagrams with $N$ rows and $k$ columns.
We discuss some generalizations of this configuration in M-theory carrying angular momentum, some of them with an interpretation as giant gravitons. We provide the microscopical description that allows to explore the region of finite 't Hooft coupling.\\ 
\\

\newpage

\tableofcontents

\def\baselinestretch{1.2}
\baselineskip 20 pt

\section{Introduction}

Giant gravitons \cite{GST} have proven to be very useful in the context of the AdS/CFT correspondence in matching D-brane configurations in string/M theory with gauge-invariant operators in the dual gauge theory. In particular, the fact that $\frac12$ BPS giant gravitons are dual to 
Schur polynomials \cite{BBNS,CJR,Beren} have helped elucidating that many properties such as the stringy exclusion principle of \cite{MS} or details about the global and local geometries of the dual branes are encoded in the gauge theory \cite{BBFH,Mello,Mello2}.

In the context of the more recent duality between Type IIA in $AdS_4\times CP^3$ and the 3-dimensional ${\cal N}=6$ Chern-Simons-matter theory with gauge group $U(N)_k\times U(N)_{-k}$ known as the ABJM theory \cite{ABJM}, giant graviton configurations in both the Type IIA and gauge theory sides have been studied in \cite{BT,NT,BP,HMPS,HMP,Dey}. The Penrose limit has been analyzed in \cite{NT2}.

In the gravity side of the Type IIA/ABJM correspondence a D2-brane spherical giant graviton with analogous properties\footnote{See however \cite{HMPS}.} to the usual dual giant gravitons in $AdS_m \times S^n$ backgrounds \cite{GMT,HHI} was constructed in \cite{NT,HMPS}. The corresponding field theory dual was worked out in \cite{SS,BP}.
A new dual graviton D2-brane solution specific to the $AdS_4\times CP^3$ background was also constructed in \cite{NT}. This D2-brane was obtained by taking the orbifold reduction of a spinning dual giant graviton in $AdS_4\times S^7/\mathbb{Z}_k$, constructed by generalizing the usual dual giant graviton ansatz to include a winding number along an angular direction in $AdS_4$. The angular momentum leads to D0-brane charge after the reduction, whereas the winding number gives F-string charge. 
The spinning D2-brane can then be regarded as a bound state of a D2-brane, D0-branes and F-strings. Since the D0-branes and F-strings generate magnetic and electric flux, a Poynting vector is generated that produces an angular momentum. This spinning D2-brane preserves some fraction of the supersymmetry, has finite energy for a  
toroidal topology, and for large angular momentum approaches a ring-like object. The identification of the dual BPS operator in this case is a difficult task, since it should not only encode Gauss' law \cite{BBFH} but also the non-zero genus. Some steps in this direction were taken in \cite{BP} (see also \cite{KKL}).
 
Supersymmetric states in the dual ABJM theory have also been predicted by reducing on the orbifold giant graviton solutions in $AdS_4\times S^7/\mathbb{Z}_k$. One such example is the spherical D2-brane with D0-charge contained in the $AdS_4$ part of $AdS_4\times CP^3$, constructed in \cite{NT}. In this paper we work out another example by reducing on the orbifold the spherical M5-brane giant graviton expanding in $S^7/\mathbb{Z}_k$. This solution gives rise to a supersymmetric static NS5-brane wrapping a twisted 5-sphere with D0-brane charge. In the maximal case the energy of the solution can be accounted for both by a bound state of $k$ D4-branes wrapping the $CP^2$, which suggests a realization in the dual field theory in terms of $k$ dibaryons, and $N$ D0-branes, which suggests a field theory realization in terms of $N$ 't Hooft monopoles. This is in agreement with the field theory, where we need to consider representations labelled by Young diagrams with $N$ rows and $k$ columns, where a single row gives a D0-brane and a single column a D4-brane \cite{ABJM}. Further, it was pointed out in \cite{ABJM} that this instability could be realized in the string theory side in terms of a NS5-brane instanton that turns the $k$ D4-branes into $N$ D0-branes. Our NS5-brane construction provides an explicit realization of this idea. In the non-maximal case the field theory dual to the NS5-brane with D0-charge should be given in terms of smaller subdeterminants  \cite{BT}.

The paper is organized as follows. In section 2 we describe the 5-sphere giant graviton in $AdS_4\times S^7/\mathbb{Z}_k$ using the action for an M5-brane wrapped on an isometric direction given in \cite{JLR}. Given that the M2-branes that end on this brane must also be wrapped  on the isometric direction the action does not contain a self-dual worldvolume 2-form but a vector field, and a closed form can be given. For an M5-brane with the topology of an $S^5$ the isometric direction is the 
coordinate along the fibre in the decomposition of the $S^5$ as a $U(1)$ fibre over the $CP^2$.
We provide a generalization of this construction by inducing further angular momenta and by taking the M5-brane wrapped instead on the $S^1/\mathbb{Z}_k$ orbifold direction, this with an aim at getting giant graviton solutions in the reduction to Type IIA. We show however that only the maximal case can be  given an interpretation in terms of giant gravitons.
Together with this so-called macroscopical description, in terms of spherical M5-branes with momentum charge, we provide in section 3 the complementary microscopical description in terms of M-theory gravitons expanding into fuzzy submanifolds of $S^7/\mathbb{Z}_k$ due to Myers dielectric effect \cite{Myers}. In Type IIA this description allows to explore the region of finite 't Hooft coupling.
In section 4 we dimensionally reduce the M5-brane giant graviton solution to produce a static NS5-brane expanding into a twisted 5-sphere inside the $CP^3$ with D0-brane charge. This configuration is described macroscopically in terms of the action describing wrapped NS5-branes in Type IIA. We also show in this section that the reduction of the M5-brane wrapped on the $S^1/\mathbb{Z}_k$ direction gives rise to a D4-brane wrapped on a deformed $CP^2$ with momentum charge which cannot however be interpreted as a giant graviton away from the maximal case.
In section 5 we provide the complementary description of the NS5-brane configuration in terms of dielectric Type IIA gravitons, and sketch the description of the D4-brane wrapped on the $CP^2$ in terms of expanding D0-branes with angular momentum.
We discuss these results and future directions in section 6.

\section{Giant gravitons in $AdS_4\times S^7/\mathbb{Z}_k$}

In this section we construct the M5-brane giant graviton solution in $AdS_4\times S^7/\mathbb{Z}_k$ using the action for a wrapped M5-brane given in \cite{JLR}. In this action the direction in which the M5-brane is wrapped occurs as a special isometric direction, so its use is limited to backgrounds with Abelian isometries. In our particular example we take the M5-brane wrapped on the
$S^5\subset S^7/\mathbb{Z}_k$ and propagating along the $S^1/\mathbb{Z}_k$ direction. The isometry is then that associated to translations along the fibre in the decomposition of the $S^5$ as a $U(1)$ fibration over the $CP^2$. In the last section we interchange the role played by the $S^1$ and $S^1/\mathbb{Z}_k$ directions, and show that the resulting configuration cannot be interpreted as a giant graviton away from the maximal case.

We start by collecting some useful formulae of the $AdS_4\times S^7/\mathbb{Z}_k$ background.

\subsection{The background}

In our conventions the $AdS_4 \times S^7/{\mathbb{Z}_k}$ metric reads:
\begin{equation}
\label{M5metric}
ds^2=-(1+\frac{r^2}{4R^2})dt^2+\frac{dr^2}{1+\frac{r^2}{4R^2}}+r^2d\Omega_2^2+R^2 ds^2_{S^7/{\mathbb{Z}_k}}
\end{equation}
with $R$ the radius of curvature in Planck units,
\begin{equation}
R=(32\pi^2 Nk)^{1/6}
\end{equation}
This is a good description of the gravity dual of the $U(N)_k\times U(N)_{-k}$ CS-matter theory of \cite{ABJM} when $N>>k^{1/5}$.
Writing the $S^7/{\mathbb{Z}_k}$ metric in coordinates adapted to its decomposition as an $S^1/{\mathbb{Z}_k}$ bundle over the $CP^3$ we have
\begin{equation}
ds^2_{S^7/{\mathbb{Z}_k}}=(\frac{1}{k}d\tau+\mathcal{A})^2+ds^2_{CP^3}
\end{equation}
where $\tau\in[0,2\pi]$. The metric of the $CP^3$ can in turn be written as (e.g. \cite{PW})
\begin{eqnarray}
\label{theCP^3}
ds_{CP^3}^2=&&d\mu^2+\sin^2\mu\,\Big[ d\alpha^2+\frac{1}{4}\sin^2\alpha\, \big(\cos^2\alpha\,(d\psi-\cos\theta\, d\phi)^2+d\theta^2+\sin^2\theta\, d\phi^2\big)\nonumber \\
&&+\frac{1}{4}\cos^2\mu\, \big(d\chi+\sin^2\alpha\, (d\psi-\cos\theta\, d\phi)\big)^2\Big]
\end{eqnarray}
where
\begin{equation}
0\le \mu,\,\alpha\le \frac{\pi}{2}\, ,\quad 0\le \theta \le \pi\, ,\quad 0\le \phi\le 2\pi\, ,\quad 0\le \psi,\, \chi\le 4\pi\ ,
\end{equation}
while the connection $A$ reads
\begin{equation}
\mathcal{A}=\frac{1}{2}\sin^2\mu\, \Big(d\chi+\sin^2\alpha\,\big(d\psi-\cos\theta\, d\phi)\Big)\, .
\end{equation}

Taking the ansatz $\mu={\rm constant}$ in $ds^2_{CP^3}$ and redefining $\chi\equiv \chi/2$ we find that $ds^2_{S^7/{\mathbb{Z}_k}}$ reduces to
\begin{equation}
\label{ansatz1}
ds^2_{S^7/{\mathbb{Z}_k}}=\frac{1}{k^2}d\tau^2+\frac{2}{k}\sin^2{\mu}\, d\tau (d\chi +A)+\sin^2{\mu}\, ds^2_{S^5}
\end{equation}
where now 
\begin{equation}
\label{connection}
A=\frac12 \sin^2{\alpha}\, (d\psi-\cos{\theta}d\phi)
\end{equation}
 and the $S^5$ is written in coordinates adapted to its decomposition as an $S^1$ bundle over the $CP^2$: 
\begin{equation}
\label{elCP^2}
ds^2_{S^5}=(d\chi+A)^2+ds^2_{CP^2}\, ,
\end{equation}
where $ds^2_{CP^2}$ is the Fubini-Study metric of the $CP^2$ (e.g. \cite{Pope}):
\begin{equation}
ds^2_{CP^2}=d\alpha^2+\frac14 \sin^2{\alpha}\Bigl[\cos^2{\alpha}\, (d\psi-\cos{\theta}d\phi)^2+d\theta^2+\sin^2{\theta}\, d\phi^2\Bigr]
\end{equation}

The 6-form potential of the $AdS_4\times S^7/\mathbb{Z}_k$ background reads, in turn:
\begin{equation}
\label{6form}
C_6=\frac{R^6}{k}\sin^6{\mu}\,d\chi\wedge d\tau\wedge {\rm dVol}(CP^2)
\end{equation}

\subsection{The 5-sphere giant graviton}

Let us now take an M5-brane wrapping the $S^5$ with radius $R\sin{\mu}$ in (\ref{ansatz1}), located at $r=0$ in $AdS_4$, and propagating on the $S^1/\mathbb{Z}_k$ fibre direction, $\tau$:
\begin{equation}
\label{theM5}
ds^2=-dt^2+R^2\Bigl[\frac{1}{k^2}d\tau^2+\frac{2}{k}\sin^2{\mu}\, d\tau (d\chi+A)+\sin^2{\mu}\, ds^2_{S^5}\Bigr]
\end{equation}

Note that $\chi$ is an isometric direction, parameterizing the $S^1$ bundle of the $S^5$. Therefore, we can use the action constructed in \cite{JLR} in order to study the M5-brane in this background. This action was successfully used in the description of the 5-sphere giant \cite{GST} and dual giant graviton \cite{GMT,HHI} solutions in $AdS_4\times S^7$ and $AdS_7\times S^4$, respectively.
The action for the Type IIA D4-brane arises from this action when reducing along the isometric direction. 

We can describe a wrapped M5-brane in the $AdS_4\times S^7/\mathbb{Z}_k$ background with the action \cite{JLR}:
\begin{equation}
\label{theM5action}
S=T_4 \int d^5\xi\, \Bigl\{- k \sqrt{|{\rm det} (P[{\cal G}] +\frac{2\pi}{k}{\cal F})|}+P[i_k C_6]+\frac12 (2\pi)^2  P[k^{-2} k_1]\wedge {\cal F}\wedge {\cal F}\Bigr\}
\end{equation}
Here $k^\mu$ is an Abelian Killing vector that points on the isometric $U(1)$ direction, ${\cal G}$ is the reduced metric ${\cal G}_{\mu\nu}=g_{\mu\nu}-k^{-2}k_\mu k_\nu$ and $i_k C_p$ denotes the interior product of the $C_p$ potential with the Killing vector. The action is therefore manifestly isometric under translations along the Killing direction. 
${\cal F}$ is the field strength associated to M2-branes, wrapped on the isometric direction, ending on the M5-brane: ${\cal F}=F+\frac{1}{2\pi}P[i_k C_3]$. When reducing along the isometric direction it gives the BI field strength of the D4-brane. We have denoted $T_4$ the tension of the brane to explicitly take into account that its spatial worldvolume is effectively 4-dimensional.

$k^{-2}k_1$ is, explicitly, $g_{\mu \chi}/g_{\chi\chi}\, dx^\mu$, in coordinates adapted to the isometry, $k^\mu=\delta^\mu_{\chi}$, and is therefore identified as the momentum operator in the isometric direction. Momentum along this direction can then be turned on by a convenient choice of ${\cal F}$, such that $\int {\cal F}\wedge {\cal F}$ is non-vanishing. Since our giant graviton solution will only propagate along the $S^1/\mathbb{Z}_k$ direction we will for the moment switch ${\cal F}$ to zero. 

Note that the 6-form potential of the $AdS_4\times S^7/\mathbb{Z}_k$ background couples in the action through
\begin{equation}
P[i_k C_6]=C^{(6)}_{\chi\tau\alpha_1\dots\alpha_4}\dot{\tau}
\end{equation}
where $\alpha_i$, $i=1,\dots, 4$ span the $CP^2$ directions.

Substituting the background fields in (\ref{theM5action}) and integrating over the $CP^2$ we find:
\begin{equation}
\label{actionM5}
S=\int dt \Bigl[-\frac{Nk}{R}\sin^5{\mu}\, \sqrt{1-\frac{R^2}{k^2}\cos^2{\mu}\, \dot{\tau}^2}+N\sin^6{\mu}\, \dot{\tau}\Bigr]
\end{equation}
where we have taken units in which the tension of a 0-brane is equal to one.

The Hamiltonian reads, in turn:
\begin{equation}
\label{hamM5}
H=\frac{k}{R}\, P_\tau \sqrt{1+\tan^2{\mu}\Bigl(1-\frac{N}{P_\tau}\sin^4{\mu} \Bigr)^2}
\end{equation} 
in terms of the conserved $\tau$ conjugate momentum, $P_\tau$. Clearly,  the minimum energy solution is reached when $\mu=0$ or
\begin{equation}
\sin{\mu}=\Bigl(\frac{P_\tau}{N}\Bigr)^{1/4}
\end{equation}
In both cases the BPS bound 
\begin{equation}
E=\frac{k}{R}P_\tau
\end{equation}
is reached, and therefore the two solutions correspond, respectively, to the point-like and giant gravitons \cite{GST} of the $AdS_4\times S^7/\mathbb{Z}_k$ background. The giant graviton solution satisfies that $P_\tau\le N$, as in \cite{GST}. Therefore the bound for the angular momentum depends only on the rank of the gauge group, in agreement with the stringy exclusion principle of \cite{MS}. The size of the giant graviton is maximal when $\mu=\pi/2$, for which the angular momentum reaches its maximum value $P_\tau=N$, and $E=k\,N/R$. 
The full energy of the maximal giant graviton can then be accounted for either by a bound state of $P_\tau=N$ gravitons (or M0-branes), with energy $k/R$, or by a bound state of $k$ M5-branes wrapped on the $S^5$, with energy $N/R$. Since in the quotient space the $k$ M5-branes reduce to $k$ D4-branes wrapping the $CP^2$ the maximal giant is dual to $k$ dibaryons \cite{ABJM,BT,BP}.

As we will see in detail in section 4, the giant graviton M5-brane is realized in Type IIA as an NS5-brane wrapping a twisted 5-sphere inside the $CP^3$. This NS5-brane is motionless, since the momentum along the M-theory circle gets replaced by D0-brane charge. The stringy exclusion principle of \cite{MS} is then realized in the $AdS_4\times CP^3$ background in terms of giant D0-branes expanded in a twisted NS5-brane.

From the previous discussion it is clear that if we want to obtain a moving brane in Type IIA we need to induce momentum in the eleven dimensional configuration in a direction different from the eleventh direction. In the next section we generalize the previous M5-brane construction to include momentum along the isometric worldvolume direction. Although as we will see the resulting configuration does not behave as a giant graviton away from the maximal case it will be useful in section 3 in order to identify the microscopical set-up that will allow to describe the expanded M5-brane in terms of dielectric gravitons.

\subsection{The 5-sphere giant graviton with magnetic flux}

Let us now generalize the giant graviton solution constructed before to include a magnetic flux  inducing momentum along the isometric direction. 

As we discussed in the previous section momentum along this direction can be switched on by a convenient choice of magnetic flux, such that
\begin{equation}
\frac{T_4}{2}(2\pi)^2\int_{\mathbb{R}\times CP^2} P[k^{-2}k_1]\wedge {\cal F}\wedge {\cal F}=P_\chi \int_{\mathbb{R}} P[k^{-2}k_1]
\end{equation}
This is achieved with $F=2\sqrt{2P_\chi}\, J$, with $J$ the K\"ahler form of the $CP^2$, $J=\frac12 dA$. The action (\ref{actionM5}) is then modified according to
\begin{equation}
\label{actionM52}
S=\int dt \Bigl[-\frac{1}{R\sin{\mu}}\Bigl(Nk\sin^6{\mu}+P_\chi\Bigr)\sqrt{1-\frac{R^2}{k^2}\cos^2{\mu}\, \dot{\tau}^2}+\frac{1}{k}\Bigl(Nk\sin^6{\mu}+P_\chi\Bigr)\, \dot{\tau}\Bigr]
\end{equation}
and the new Hamiltonian reads
\begin{equation}
\label{hamM52}
H=\frac{k}{R}\, P_\tau \sqrt{1+\tan^2{\mu}\Bigl(1-\frac{Nk\sin^6{\mu}+P_\chi}{kP_\tau \sin^2{\mu}}\Bigr)^2}\, .
\end{equation} 
Written in this way it is clear that $H(\mu)$ is minimum for $\mu=0$ and $\sin{\mu}$ satisfying
\begin{equation}
\label{M5general}
Nk\sin^6{\mu}-kP_\tau\sin^2{\mu}+P_\chi=0\, ,
\end{equation}
and that for both these point-like and expanded brane solutions 
\begin{equation}
H=\frac{k}{R}\, P_\tau
\end{equation}
Since for $P_\chi\ne 0$ the energy depends only on one of the two conserved charges the BPS bound is not satisfied. Nonetheless
in the maximal case the $S^1/\mathbb{Z}_k$ and $S^1$ directions become parallel and we have from  (\ref{M5general}) that $P_\tau=N+P_\chi/k$, so in this case there is only one independent conserved charge. A non-vanishing $P_\chi$  allows then the maximal giant to propagate with an arbitrary angular momentum on the $S^1/\mathbb{Z}_k$ direction.

In Type IIA language this expanded solution is realized in terms of a NS5-brane wrapping a twisted 5-sphere submanifold of the $CP^3$, with D0-brane charge $P_\tau$ and momentum $P_\chi$. As before, the energy only depends on the D0-brane charge, and therefore away from the maximal case the configuration is not BPS. Moreover, it does not have an interpretation as a giant graviton. However, inspired by this result we can think of  interchanging the role played by the $S^1/\mathbb{Z}_k$ and $S^1$ directions. Namely, we can take the M5-brane wrapping the submanifold of $S^7/\mathbb{Z}_k$ spanned by $\tau$ and the coordinates parameterizing the $CP^2$, and propagating on $\chi$. This configuration becomes a D4-brane wrapped on a ``squashed'' $CP^2$ and propagating on $\chi$ in Type IIA. We will see however that it does not have an interpretation as a giant graviton away from the maximal case.

\subsection{The giant graviton wrapped on the $S^1/\mathbb{Z}_k$ direction}

As we have just mentioned we can similarly consider an M5-brane\footnote{In our conventions we need to consider in fact an anti-M5-brane with opposite $\tau$-momentum.} wrapped on the 
submanifold of $S^7/Z_k$ with metric
\begin{equation}
\label{M5twisted}
ds^2=R^2\Bigl[\frac{1}{k^2}d\tau^2+\frac{2}{k}\sin^2{\mu}\, A\, d\tau +\sin^2{\mu}\, (ds^2_{CP^2}+A^2)\Bigr]
\end{equation}
and propagating on the $\chi$ direction. 
Given that $\tau$ is an isometric direction we can still use the action (\ref{theM5action}) with $k^\mu=\delta^\mu_\tau$. 
For the sake of generality we also switch a magnetic flux 
$F=2\sqrt{2P_\tau}\, J$ on, inducing $P_\tau$ momentum through the coupling
\begin{equation}
\frac{T_4}{2}(2\pi)^2\int_{\mathbb{R}\times M_4} P[k^{-2}k_1]\wedge {\cal F}\wedge {\cal F}
\end{equation}
The configuration simplifies a lot if we also
induce an electric flux  proportional to the connection on the $CP^2$: $E_i=R\sin{\mu}\cos{\mu}\, A_i$. Then, substituting in the action (\ref{theM5action}) we find:
\begin{equation}
S=\int dt \Bigl[ -\frac{1}{R}\Bigl(N\sin^4{\mu}+kP_\tau\Bigr)\sqrt{1-R^2\sin^2{\mu}\cos^2{\mu}\, \dot{\chi}^2}\,+\, \sin^2{\mu}\Bigl(N\sin^4{\mu}+kP_\tau\Bigr)\dot{\chi}\Bigr]
\end{equation}
and a Hamiltonian
\begin{equation}
\label{hamM5twisted}
H=\frac{P_\chi}{R\sin{\mu}}\sqrt{1+\tan^2{\mu}\Bigl(1-\frac{N\sin^4{\mu}+kP_\tau}{P_\chi}\Bigr)^2}
\end{equation}
This expression is very similar to the Hamiltonian (\ref{hamM52}) describing the spherical M5-brane with magnetic flux. However in this case one can easily see that there are no solutions for which $E=P_\chi/R$ unless $\mu=\pi/2$, which leads us back to the maximal case. In this case the M5-brane wraps a $S^5/\mathbb{Z}_k$ submanifold of $S^7/\mathbb{Z}_k$ and the induced electric flux vanishes. Also $P_\chi=N +k P_\tau$, and the brane is allowed to move with an arbitrary angular momentum in the $S^1$ direction. For vanishing $P_\tau$ we recover the usual maximal giant graviton with momentum reaching the bound imposed by the stringy exclusion principle, wrapped in this case on a $S^5/\mathbb{Z}_k$ submanifold. In the reduction to Type IIA the $S^1$ direction shrinks to a point, and the $N$ charge can no longer be interpreted as momentum charge. The configuration is instead interpreted as a dibaryon with energy $N/L$. Our analysis shows that the dibaryon arises in the gravity side as the limiting case of a D4-brane wrapping the ``squashed'' $CP^2$ included in
\begin{equation}
ds^2=-dt^2+L^2\sin^2{\mu}\, \Bigl[\cos^2{\mu}\, (d\chi+A)^2+ds^2_{CP^2}\Bigr]
\end{equation}
and propagating along the $\chi$ direction.

\section{A description in terms of expanding gravitons}

In this section we show that the previous configurations can alternatively be described in terms of gravitons expanding into fuzzy M5-branes.  This is the microscopical realization of the macroscopical M5-branes wrapping classical submanifolds of $S^7/\mathbb{Z}_k$ with angular momenta. This description is valid in the supergravity limit $R>>1$, whereas the microscopical description is good when the mutual separation of the expanding gravitons is much smaller than the string length. For $\cal N$ expanding gravitons this is fixed by the condition $R<<{\cal N}^{1/4}$. The two descriptions are then complementary for finite ${\cal N}$ and should agree in the large ${\cal N}$ limit, where they have a common range of validity. In $AdS_4\times CP^3$ the regime of validity of the microscopical description is fixed by the condition that $N<<{\cal N} k$. Therefore this description allows to explore the region of finite 't Hooft coupling. We will see that the same conclusions regarding the existence of giant graviton configurations that we reached in the macroscopical set-up will hold microscopically.

\subsection{The action for M-theory gravitons}

We start from the action for coincident gravitons, or gravitational waves, in M-theory constructed in \cite{JL}. Using this action it was possible to describe microscopically the giant and dual giant graviton solutions in $AdS_4\times S^7$ and $AdS_7\times S^4$ \cite{JL,JLR}, and to derive
Matrix theory in the maximally supersymmetric pp-wave background of M-theory \cite{BMN} with an extra non-perturbative coupling giving rise to the so-far elusive 5-sphere giant graviton \cite{LR}. 
The BI part of the action is given by
\begin{equation}
\label{DBIM0}
S_{DBI}=-\int d\xi^0\, {\rm STr}\Bigl\{k^{-1}\sqrt{|P[E_{\mu\nu}+E_{\mu i}(Q^{-1}-\delta)^i_k E^{kj}E_{j\nu}]{\rm det}Q|}\Bigr\}
\end{equation}
where $E_{\mu\nu}={\cal G}_{\mu\nu}+k^{-1}(i_k C_3)_{\mu\nu}$, ${\cal G}$ is the reduced metric defined in section 2, 
$\mu, \nu$ denote spacetime indices and $i,j$ spatial ones, and $Q$ is given by
$Q^i_j=\delta^i_j+\frac{ik}{2\pi}[X^i,X^k]E_{kj}$.
The CS action reads
\begin{equation}
\label{CSM0}
S_{CS}=\int d\xi^0\,{\rm STr}\Bigl\{P[k^{-2}k_1]+\frac{i}{2\pi}P[(i_X i_X)C_3]-\frac12\frac{1}{(2\pi)^2}P[(i_X i_X)^2 i_k C_6]+\dots\Bigr\}
\end{equation}
In this action $k^\mu$ is an Abelian Killing vector that points on the direction of propagation of the waves. As in action (\ref{theM5action}) this direction is isometric because the background fields are either contracted with the Killing vector or pulled-back in the worldvolume with covariant derivatives relative to the isometry: 
\begin{equation}
\label{covder}
{\cal D}_0 X^\mu=\partial_0 X^\mu-k^{-2}k_\nu\, \partial_0X^\nu k^\mu
\end{equation}
In this way the dependence on the isometric direction is effectively eliminated from the action. This action is in fact a gauge fixed action in which the $U({\cal N})$ vector field, associated to M2-branes (wrapped on the direction of propagation) ending on the waves, has been taken to vanish. In this gauge $U({\cal N})$ covariant derivatives reduce to ordinary derivatives, and gauge covariant derivatives can be defined using ordinary derivatives as in (\ref{covder}).

The action given by (\ref{DBIM0}) and (\ref{CSM0}) was constructed in \cite{JL} by uplifting to eleven dimensions the action for Type IIA gravitational waves derived in \cite{JL2} using Matrix String theory in a weakly curved background, and then going beyond the weakly curved background approximation by demanding agreement with Myers action for D0-branes when the waves propagate along the eleventh direction. In the action for Type IIA waves the circle in which Matrix theory is compactified in order to construct Matrix String theory cannot be decompactified in the non-Abelian case \cite{JL2}. In fact, the action exhibits a $U(1)$ isometry associated to translations along this direction, which by construction is also the direction on which the waves propagate. A simple way to see this is to recall that the last operation in the 9-11 flip involved in the construction of Matrix String theory is a T-duality from fundamental strings wound around the 9th direction. Accordingly, in the action we find a minimal coupling to $g_{\mu 9}/g_{99}$, which is the momentum operator $k^{-2}k_1$ in adapted coordinates. Therefore, by construction, the sum of the actions (\ref{DBIM0}) and (\ref{CSM0}) is designed to describe BPS waves with momentum charge along the compact isometric direction. It is important to mention that in the Abelian limit, when all dielectric couplings and $U({\cal N})$ covariant derivatives disappear, the action can be Legendre transformed into an action in which the dependence on the isometric direction has been restored. This action is precisely the usual action for a massless particle written in terms of an auxiliary $\gamma$ metric (see \cite{JL} and \cite{JL2} for the details), where no information remains about the momentum charge carried by the particle.

\subsection{Expanded graviton configurations}

Microscopically the spherical M5-brane described in the previous section is built up of gravitons expanding into a fuzzy 5-sphere through Myers dielectric effect. In our construction the fuzzy $S^5$ will simply be defined as an $S^1$ bundle over a fuzzy $CP^2$. As we will see the existence of the $S^1$ direction is crucial in order to find the right dielectric coupling that will cause the expansion of the gravitons. This construction of the fuzzy 5-sphere has been successfully used in the microscopical description of various configurations involving 5-spheres (see for instance \cite{JLR, LR,JLR2, JLR3}). In all cases it brings back the right macroscopical description when the number of constituents is large. 

The calculation in this section is very similar to the microscopical description of giant gravitons expanding in $AdS_4\times S^7$ presented in \cite{JLR}. 

The fuzzy $CP^2$ has been extensively studied in the literature in various contexts. In the context of Myers dielectric effect it was first studied in \cite{TV} and then in \cite{JLR, LR,JLR2, JLR3, LPSS}. 
In general $G/H$ coset manifolds can be described as fuzzy surfaces if $H$ is the isotropy group of the lowest weight state of a given irreducible representation of G \cite{Madore,TV}. Since different irreducible representations have associated different isotropy groups they can give rise to different cosets $G/H$. $CP^2$ has $G=SU(3)$, $H=U(2)$, and this is precisely the isotropy group of the $SU(3)$ irreducible representations $(m,0)$, $(0,m)$, where we parameterize the irreducible representations of $SU(3)$ by two integers $(m,m^\prime)$ corresponding to the number of fundamental and anti-fundamental indices. One can also take a more geometrical view more suitable for our purposes. Using the fact that $CP^n$ spaces can be defined as the submanifolds of $\mathbb{R}^{n^2+2n}$ determined by a given set of $n^2$ constraints, a fuzzy version arises by promoting the Cartesian coordinates that embed the $CP^n$ in $\mathbb{R}^{n^2+2n}$ to $SU(n+1)$ matrices in the irreducible totally symmetric representations $(m,0)$ or $(0,m)$. Indeed only for these representations can the set of $n^2$ constraints be realized at the level of matrices. The Cartesian coordinates are then taken to play the role of the non-Abelian transverse scalars that couple in the action for coincident gravitons. 

For the $CP^2$ the 8 Cartesian coordinates that embed it in $\mathbb{R}^8$ satisfy
\begin{equation}
\sum_{i=1}^{8}x^i x^i=1\, \qquad \sum_{j,k=1}^{8} d^{ijk} x^j x^k=\frac{1}{\sqrt{3}}x^i
\end{equation}
where $d^{ijk}$ are the components of the totally symmetric $SU(3)$-invariant tensor. In these coordinates the Fubini-Study metric is given by
\begin{equation}
ds^2_{CP^2}=\frac13 \sum_{i=1}^{8} (dx^i)^2
\end{equation}
This set of constraints can be implemented at the level of matrices if we choose the set of coordinates $X^i$ $(i=1,\dots 8)$ in the irreducible totally symmetric representation of order $m$, $(m,0)$, satisfying
\begin{equation}
[X^i,X^j]=i\Lambda_{(m)}f^{ijk}X^k\, , \qquad \Lambda_{(m)}=\frac{1}{\sqrt{\frac{m^2}{3}+m}}
\end{equation}
with $f^{ijk}$ the structure constants of $SU(3)$, $[\lambda^i. \lambda^j]=2if^{ijk}\lambda^k$.
The dimension of the $(m,0)$ representation is given by
\begin{equation}
{\cal N}=\frac{(m+2)(m+1)}{2}
\end{equation}

In the $AdS_4\times S^7/\mathbb{Z}_k$ background we take the gravitons located in $r=0$ and expanding into the fuzzy $S^5$ with radius $R\sin{\theta}$ inside $S^7/\mathbb{Z}_k$ described by the metric (\ref{theM5}), in which we take Cartesian coordinates to parameterize the $CP^2$.
We choose $k^\mu=\delta^\mu_\chi$, so that the gravitons carry by construction $P_\chi$ momentum, and take 
$\tau=\tau(t)$ in order to induce $P_\tau$ momentum. We then have that
\begin{eqnarray}
&&k=R\sin{\mu}\, ,\qquad E_{00}=-1+\frac{R^2}{k^2}\cos^2{\mu}\, {\dot{\tau}}^2\, , \nonumber\\
&&Q^i_j=\delta^i_j-\frac{R^3\sin^3{\mu}}{2\pi\sqrt{m^2+3m}}f_{ijk}X^k\, , \qquad i,j=1,\dots,8\, .
\end{eqnarray}
The determinant of $Q$ can be computed as explained for instance in \cite{LPSS}, with the result
\begin{equation}
{\rm det}Q=\Bigl(1+\frac{R^6\sin^6{\mu}}{16\pi^2(m^2+3m)}\Bigr)^2\mathbb{I}
\end{equation}
As explained in \cite{LPSS} this expression is valid in the limit $R>>1$, $m>>1$, with $R^3/m$ finite.

Substituting in the DBI action (\ref{DBIM0}) we find
\begin{equation}
S_{DBI}=-\frac{{\cal N}}{R\sin{\mu}}\Bigl(1+\frac{R^6\sin^6{\mu}}{16\pi^2(m^2+3m)}\Bigr)\int dt\, \sqrt{1-\frac{R^2}{k^2}\cos^2{\mu}\,{\dot{\tau}}^2}
\end{equation}
where ${\cal N}$ arises as ${\rm dim}(m,0)={\rm STr}\, \mathbb{I}$.
Using that
\begin{equation}
C_6=\frac{R^6\sin^6{\mu}}{2k}d\chi\wedge d\tau\wedge J\wedge J
\end{equation}
and that (see \cite{LPSS})
\begin{equation}
J_{ij}=\frac{1}{3\sqrt{3}}f_{ijk}X^k
\end{equation}
we find the CS action:
\begin{equation}
S_{CS}=\int dt\, \frac{{\cal N}}{k}\Bigl(1+\frac{R^6\sin^6{\mu}}{16\pi^2(m^2+3m)}\Bigr)\dot{\tau}
\end{equation}
In terms of the conserved conjugate momentum to $\tau$ we have the Hamiltonian
\begin{equation}
\label{hmic}
H=\frac{k}{R}P_\tau\sqrt{1+\tan^2{\mu}\Bigl(1-\frac{{\cal N}}{kP_\tau\sin^2{\mu}}(1+\frac{2Nk\sin^6{\mu}}{m^2+3m})\Bigr)^2}
\end{equation}

Taking into account that the $P_\chi$ momentum of the configuration is by definition the number of gravitons ${\cal N}$, in our units in which the tension is set to one, we can rewrite (\ref{hmic}) as
\begin{equation}
\label{hmic2}
H=\frac{k}{R}P_\tau\sqrt{1+\tan^2{\mu}\Bigl(1-\frac{1}{kP_\tau\sin^2{\mu}}(P_\chi+\frac{2{\cal N}}{m^2+3m}Nk\sin^6{\mu})\Bigr)^2}
\end{equation}
from which it is clear that both Hamiltonians (\ref{hmic2}) and (\ref{hamM52}) exactly agree in the large $m$ limit, where ${\cal N}\sim m^2/2$. Moreover, in the limit $P_{\chi}\rightarrow 0$ we can also describe the giant graviton configuration of section 2.2, realized as a spherical M5-brane with just $P_\tau$ momentum, with the Hamiltonian
\begin{equation}
H=\frac{k}{R}P_\tau\sqrt{1+\tan^2{\mu}\Bigl(1-\frac{2{\cal N}}{m^2+3m}\frac{N\sin^4{\mu}}{P_\tau}\Bigr)^2}\, ,
\end{equation}
which agrees exactly with the Hamiltonian (\ref{hamM5}) in the large $m$ limit. Note that the difference between $P_\chi$ being zero or not is merely a coordinate transformation, a boost in $\chi$. How to perform coordinate transformations in non-Abelian actions is however an open problem \cite{dBS,dBS2,BFLvR,BKLvR,AJT}. In this case the way in which the limit $P_\chi\rightarrow 0$ should be taken is dictated by the agreement with the macroscopical description.

Finally, we can briefly sketch the microscopical description of the $S^5/\mathbb{Z}_k$ maximal giant graviton of section 2.4. In this case we take $k^\mu=\delta^\mu_\tau$,  so that the gravitons are wrapped on $\tau$ and carry by construction $P_\tau$ momentum, and take $\chi=\chi(t)$ in order to induce $P_\chi$ momentum. The fuzzy $S^5/\mathbb{Z}_k$ on which the gravitons expand is then defined as an $S^1/\mathbb{Z}_k$ fibre over a fuzzy $CP^2$. Substituting in (\ref{DBIM0}), (\ref{CSM0}) we find the Lagrangian
\begin{equation}
L=k{\cal N}(-\frac{1}{R}+\dot{\chi})(1+\frac{2N}{k(m^2+3m)})
\end{equation}
$P_\chi$ is simply given by $P_\chi=k{\cal N}+\frac{2N{\cal N}}{m^2+3m}$ and $H=\frac{P_\chi}{R}$. $P_\chi$ gives in the large $m$ limit $P_\chi=k{\cal N}+N=kP_\tau +N$, in agreement with the result in section 2.4.

\section{Dielectric branes in $AdS_4\times CP^3$}

In this section we give the Type IIA description of the 5-sphere giant graviton solution of section 2.2. We restrict to the zero flux case since we have already seen that away from the maximal case introducing flux ($\leftrightarrow P_\chi$ momentum) does not allow the construction of more general giant graviton solutions. 
In Type IIA the 5-sphere giant graviton gives rise to a NS5-brane expanding into a twisted 5-sphere inside the $CP^3$. Momentum along the M-theory circle gets replaced by D0-brane charge, so the NS5-brane is motionless. The energy of the ground state is then accounted for by a bound state of $P_\tau$ D0-branes. In the maximal case the ground state is degenerate, and can be accounted for either by a bound state of $N$ D0-branes or by a bound state of $k$ dibaryons. This is a realization on the gravity side of the duality of Young tableaux with $N$ rows and $k$ columns 
\cite{ABJM}. The stringy exclusion principle is realized in this case in terms of giant D0-branes expanded in a twisted 5-sphere NS5-brane.
We also provide the Type IIA realization of the M5-brane wrapped on $S^1/\mathbb{Z}_k$ discussed in section 2.4. This becomes a D4-brane wrapped on a ``squashed'' $CP^2$ with angular momentum which can however only be interpreted as a giant graviton in the maximal case. Still, this configuration allows to see the $k$ dibaryons emerging as the limiting case of D4-branes with angular momentum.

\subsection{The NS5-brane with D0 charge}

The reduction of (\ref{theM5}) to Type IIA gives the metric of a twisted 5-sphere, with
$\cos{\mu}$ parameterizing the twist between the $S^1$ and the $CP^2$:
\begin{equation}
\label{NS5}
ds^2=-dt^2+L^2\sin^2{\mu}\, \Bigl[\cos^2{\mu}\, (d\chi+A)^2+ds^2_{CP^2}\Bigr]\, .
\end{equation}
Here $A$ is given by (\ref{connection}) and $L$ is the radius of curvature of the $CP^3$ in string units:
\begin{equation}
L=\Bigl(\frac{32\pi^2 N}{k}\Bigr)^{1/4}
\end{equation}
There is also a RR 1-form field
\begin{equation}
\label{C1NS5}
C_1=k\sin^2{\mu}\, (d\chi + A) \, ,
\end{equation}
a 5-form potential
\begin{equation}
\label{C5NS5}
C_5=L^4 k\sin^6{\mu}\, d\chi\wedge d{\rm Vol}_{CP^2}\, 
\end{equation}
and a dilaton $e^\phi=\frac{L}{k}$.

We take the NS5-brane wrapped on the manifold with metric (\ref{NS5}). Since the background is isometric in the $\chi$ direction we can describe the NS5-brane using the action 
that arises from (\ref{theM5action}) by reducing along a transverse direction. The (bosonic) worldvolume field content of this action consists on 4 transverse scalars, a vector field, associated to D2-branes wrapped on the isometric direction, and a scalar, coming from the reduction of the eleventh direction. This scalar forms an invariant field strength with the RR 1-form potential, and is therefore associated to D0-branes. We can therefore induce D0-brane charge in the configuration through this field.
Comparing to the action of the unwrapped NS5-brane in Type IIA constructed in \cite{BLO,EJL} the self-dual 2-form field of the latter has been replaced by a vector, and this allows to give a closed form for the action.
The BI part reads:
\begin{equation}
\label{wrappedNS5}
S_{DBI}=-T_4 \int d^5\xi\, e^{-2\phi}\sqrt{k^2+e^{2\phi}(i_k C_1)^2}\sqrt{ |{\rm det}\Bigl(P[{\cal G}]+\frac{e^{2\phi}k^2}{k^2+e^{2\phi}(i_k C_1)^2}{\cal F}_1^2 \Bigr)|}
\end{equation}
Here ${\cal F}_1$ is the field strength associated to D0-branes ``ending" on the NS5-brane: ${\cal F}_1=dc_0+P[C_1]$, where the pull-back is taken with gauge covariant derivatives, as defined in 
(\ref{covder}). $c_0$ is the worldvolume scalar whose origin is the eleventh direction.
We have ignored for simplicity the contribution of the reduction of ${\cal F}$, now associated to D2-branes wrapped on $\chi$, since this field will be vanishing in our background.

The relevant part of the CS action is given by:
\begin{equation}
S_{CS}=T_4 \int i_k C_5 \wedge dc_0
\end{equation}

Substituting the background fields in the BI and CS actions and integrating over the $CP^2$ we find:
\begin{equation}
S= \int dt\Bigl[-\frac{Nk}{L} \sin^5{\mu}\sqrt{1-\frac{L^2}{k^2}\cos^2{\mu}\, {\dot c}_0^2}+N \sin^6{\mu}\,  {\dot c}_0 \Bigl]
\end{equation}
where we have taken $c_0$ time-dependent in order to induce D0-brane charge in the NS5-brane. 
This action is analogous to (\ref{actionM5}) with $L\leftrightarrow R$, $c_0\leftrightarrow \tau$. The Hamiltonian is then
\begin{equation}
\label{HNS5}
H=\frac{k}{L} M  \sqrt{1+\tan^2{\mu}\, \Bigl(1-\frac{N}{M} \sin^4{\mu}\Bigr)^2}
\end{equation}
where we have denoted with $M$ the $c_0$ conjugate momentum, which is conserved and is now interpreted as D0-brane charge. As for the giant graviton in section 2.2, the minimum energy solution is reached when $\mu=0$ or
\begin{equation}
\label{giantNS5}
\sin{\mu}=\Bigl(\frac{M}{N}\Bigr)^{1/4}
\end{equation}
In both cases $E=\frac{k}{L} M$, and we find a BPS configuration of $M$ D0-branes with energy $k/L$.
For $\mu=0$ the NS5-brane is point-like and can carry arbitrary D0-brane charge, while for $\mu$ satisfying (\ref{giantNS5}) it wraps the twisted 5-sphere described by the spatial part of the metric (\ref{NS5}) with radius $L\sin{\mu}=(32\pi^2 M/k)^{1/4}$, and $M$ has to satisfy $M\le N$. Therefore the stringy exclusion principle is realized in terms of giant D0-branes expanded into a twisted 5-sphere inside the $CP^3$. 

In the maximal case, $\mu=\pi/2$, $M=N$ and the energy can be accounted for both by a bound state of $N$ 't Hooft monopoles, with energy $k/L$, and a bound state of $k$ dibaryons, with energy $N/L$. In this case the circle of the twisted 5-sphere shrinks to a point, and the NS5-brane collapses to a D4-brane wrapping the $CP^2$. This is so because when $\mu=\pi/2$ the $\tau$ and $\chi$ directions in the eleven dimensional background become parallel, and therefore the M5-brane wrapped on $\chi$, a NS5-brane in Type IIA, and the M5-brane wrapped on $\tau$, a D4-brane, become equivalent. We show explicitly in the next section how the maximal giant in Type IIA can be realized as a D4-brane wrapping the $CP^2$ with momentum $N$. As we have already mentioned this configuration cannot however be extended beyond the maximal case to give more general D4-brane giant graviton configurations.

\subsection{A D4-brane giant graviton}

Let us now use the metric (\ref{NS5}) to describe a D4-brane wrapped on the ``squashed'' $CP^2$ with metric $g_{ij}=g_{ij}^{(CP^2)}+\cos^2{\mu}\, A_iA_j$, and propagating on the $\chi$ direction. This configuration is the IIA realization of the M5-brane wrapped on $S^1/\mathbb{Z}_k$ discussed in section 2.4. As in that case the configuration is greatly simplified if we introduce an electric flux proportional to the connection of the $CP^2$, $E_i=L\sin{\mu}\cos{\mu}\, A_i$. The role of this electric flux is to compensate the contribution of the $\cos^2{\mu}\, A_iA_j$ part of the metric to the Born-Infeld action. Since the inclusion of a magnetic flux inducing $P_\tau$ momentum in eleven dimensions did not allow the construction of more general giant graviton configurations we will take in this section $F_{ij}=0$. More general configurations with non-vanishing D0-brane charge can be considered if $F_{ij}\ne 0$ which do not have however an interpretation as giant gravitons.

Substituting in the DBI action for a single D4-brane we find
\begin{equation}
S_{DBI}=-T_4 \int d^5\xi\, e^{-\phi}\sqrt{|{\rm det}(P[g]+2\pi F)|}=-\frac{N}{L} \sin^4{\mu}\int dt \sqrt{1-L^2\sin^2{\mu}\cos^2{\mu}\dot{\chi}^2}
\end{equation}
The CS action gives in turn
\begin{equation}
S_{CS}=T_4 \int P[C_5]=N\sin^6{\mu}\int dt\, \dot{\chi}
\end{equation}
Given that $\chi$ is cyclic $P_\chi$ is conserved, and the Hamiltonian reads
\begin{equation}
\label{HD4}
H=\frac{P_\chi}{L\sin{\mu}}\sqrt{1+\tan^2{\mu}\Bigl(1-\frac{N}{P_\chi}\sin^4{\mu}\Bigr)^2}
\end{equation}
This is analogous to expression (\ref{hamM5twisted}), with $R\leftrightarrow L$ and $P_\tau=0$.
As in that case the ground state is reached when $\mu=\pi/2$, for which $P_\chi=N$ and
$H=N/L$. The $S^5/\mathbb{Z}_k$ expanded manifold reduces to a $CP^2$ and the giant is simply realized in Type IIA as a dibaryon. This analysis shows that the dibaryon arises in the gravity side as the limiting case of a D4-brane wrapping the ``squashed'' $CP^2$ included in
\begin{equation}
ds^2=-dt^2+L^2\sin^2{\mu}\, \Bigl[\cos^2{\mu}\, (d\chi+A)^2+ds^2_{CP^2}\Bigr]
\end{equation}
and propagating along the $\chi$ direction. 

\section{The microscopical description in Type IIA}

We have seen in section 3 that it is possible to describe the giant graviton configurations studied in section 2 in terms of gravitons expanding into fuzzy $S^5$ or $S^5/\mathbb{Z}_k$ manifolds inside the $S^7/\mathbb{Z}_k$ part of the eleven dimensional background. Reducing to Type IIA we find that the twisted NS5-brane with D0-brane charge is described in terms of Type IIA gravitons with D0-charge expanding into a fuzzy twisted 5-sphere inside the $CP^3$. The $S^5/\mathbb{Z}_k$ giant graviton is described in turn in terms of D0-branes with $\chi$-momentum expanding into a fuzzy $CP^2$.

The action describing coincident gravitons in Type IIA was constructed in \cite{LR2} by reducing along a transverse direction the action for M-theory gravitons reviewed in section 3. Using this action it was possible to reproduce Matrix String theory in various Type IIA pp-wave backgrounds. The BI part of the action reads:
\begin{eqnarray}
\label{IIAwaves}
S_{DBI}&=&-\int d\xi^0\, {\rm STr}\Bigl\{\frac{1}{\sqrt{k^2+e^{2\phi}(i_k C_1)^2}}\sqrt{\Bigl(\mathbb{I}-(k^2+e^{2\phi}(i_k C_1)^2)[c_0,X]^2\Bigr)\, {\rm det}Q}\nonumber\\
&&.\sqrt{|P[E]+\frac{k^2e^{2\phi}}{k^2+e^{2\phi}(i_k C_1)^2}{\cal F}_1^2|}\Bigr\}
\end{eqnarray}
where
\begin{eqnarray}
&&E_{\mu\nu}={\cal G}_{\mu\nu}+\frac{e^\phi}{\sqrt{k^2+e^{2\phi}(i_k C_1)^2}}(i_k C_3)_{\mu\nu}\nonumber\\
&&Q^i_j=\delta^i_j+i[X^i, X^k]e^{-\phi}\sqrt{k^2+e^{2\phi}(i_k C_1)^2}E_{kj} \, , \qquad i,j=1,\dots, 8
\end{eqnarray}
$c_0$ is the scalar field that comes from the reduction of the eleventh transverse direction and ${\cal F}_1$ is its invariant field strength ${\cal F}_1=dc_0+P[C_1]$, introduced in section 4.1. Therefore $c_0$ is associated to D0-branes ending on the waves. The first square root in the numerator in (\ref{IIAwaves}) comes from the reduction of the determinant of the nine dimensional $Q$ matrix, whereas the second square root comes from the reduction of the pull-back of the metric. The terms coming from the reduction of $E_{\mu i}(Q^{-1}-\delta)^i_k E^{kj} E_{j \nu}$ have been omitted from this action, since they will not contribute to the calculation in this section.

The relevant terms in the dimensional reduction of the CS action are
\begin{equation}
S_{CS}=\int d\xi^0\, {\rm STr}\Bigl\{P[k^{-2}k_1]+\frac{e^{2\phi}i_k C_1}{k^2+e^{2\phi}(i_k C_1)^2}{\cal F}_1
-\frac12\frac{1}{(2\pi)^2}P[(i_X i_X)^2i_kC_5]\wedge F\Bigr\}
\end{equation}

In the $AdS_4\times CP^3$ background we take the gravitons located in $r=0$ and expanding into the fuzzy $CP^2$ with radius $L\sin{\theta}$ contained in the $CP^3$, which we parameterize with Cartesian coordinates as in section 3. We choose $k^\mu=\delta^\mu_\chi$ and take $c_0$ commuting and time-dependent, in order to induce D0-brane charge in the configuration. We then have:
\begin{eqnarray}
&&k=L\sin{\mu}\cos{\mu}\, , \qquad i_k C_1=k\sin^2{\mu}\, , \qquad E_{00}=-1\nonumber\\
&&Q^i_j=\delta^i_j-\frac{kL^2\sin^3{\mu}}{2\pi\sqrt{m^2+3m}}f_{ijk}X^k\, , \qquad i,j=1,\dots, 8
\end{eqnarray}
Computing the determinant and substituting in the DBI action we find
\begin{equation}
S_{DBI}=-\frac{{\cal N}}{L\sin{\mu}}\Bigl(1+\frac{k^2L^4\sin^6{\mu}}{16\pi^2(m^2+3m)}\Bigr)\int dt\, \sqrt{1-\frac{L^2}{k^2}\cos^2{\mu}\,{\dot{c_0}}^2}
\end{equation}
where ${\cal N}$ arises as ${\rm dim}(m,0)={\rm STr}\, \mathbb{I}$.
The CS part reads in turn
\begin{equation}
S_{CS}=\int dt\, \frac{{\cal N}}{k}\Bigl(1+\frac{k^2 L^4\sin^6{\mu}}{16\pi^2(m^2+3m)}\Bigr)\dot{c_0}
\end{equation}
In terms of the conserved conjugate momentum to $c_0$, interpreted as D0-brane charge, $M$, we have the Hamiltonian
\begin{equation}
\label{hmicNS5}
H=\frac{k}{L}M\sqrt{1+\tan^2{\mu}\Bigl(1-\frac{{\cal N}}{kM\sin^2{\mu}}(1+\frac{2Nk\sin^6{\mu}}{m^2+3m})\Bigr)^2}
\end{equation}
As discussed in section 3 we can describe NS5-branes with only D0-brane charge taking the limit $P_\chi\rightarrow 0$, which gives
\begin{equation}
H=\frac{k}{L}M\sqrt{1+\tan^2{\mu}\Bigl(1-\frac{2N{\cal N}\sin^4{\mu}}{M(m^2+3m)}\Bigr)^2}
\end{equation}
in perfect agreememt with the Hamiltonian (\ref{HNS5}) in the large $m$ limit.

Finally, we briefly sketch the microscopical description of the D4-brane maximal giant graviton of section 4.2.
In this case we should have D0-branes with $\chi$-momentum charge expanding into a fuzzy $CP^2$. Substituting in Myers action for ${\cal N}$ coincident D0-branes we find
\begin{equation}
L=k{\cal N}(-\frac{1}{L}+\dot{\chi})(1+\frac{2N}{k(m^2+3m)})
\end{equation}
$P_\chi$ is simply given by $P_\chi=k{\cal N}+\frac{2N{\cal N}}{m^2+3m}$ and $H=\frac{P_\chi}{L}$. $P_\chi$ gives in the large $m$ limit $P_\chi=k{\cal N}+N=kM +N$, in agreement with the result in section 4.2\footnote{In section 4.2 we have taken $M=0$ but $P_\chi=kM+N$ for $M\ne 0$.}.

\section{Conclusions}

We have constructed various giant graviton configurations of M5-branes expanded in $AdS_4\times S^7/\mathbb{Z}_k$ and discussed their realization in Type IIA. 

The first configuration is an M5-brane wrapping an $S^5\subset S^7/\mathbb{Z}_k$ and propagating on the $S^1/\mathbb{Z}_k$ direction. This is the trivial extension of the giant graviton in $AdS_4\times S^7$ to the $AdS_4\times S^7/\mathbb{Z}_k$ background. This brane has been described using the action for M5-branes wrapped on an  isometric direction constructed in \cite{JLR}. Since the M2-branes that end on these branes must be wrapped on the isometric direction the self-dual 2-form is replaced by a vector field and the action admits a closed form. This configuration becomes in Type IIA a motionless NS5-brane expanding into a twisted 5-sphere inside the $CP^3$. The ground state is a BPS configuration of giant D0-branes satisfying the stringy exclusion principle of \cite{MS}. 

The second configuration that we have analyzed is the extension of the former giant graviton to include an additional momentum along the isometric direction. We have seen however that although the ground state has the energy of a giant graviton propagating on the orbifold direction it is not protected by supersymmetry. In the maximal case the configuration becomes BPS and the additional angular momentum allows the giant graviton to move with an arbitrary angular momentum. 

The last configuration that we have discussed is an M5-brane whose isometric direction points on the $S^1/\mathbb{Z}_k$ direction, propagating both on this orbifold direction and the $S^1$ fibre of the 5-sphere. The ground state corresponds to a giant graviton with maximum size wrapping a $S^5/\mathbb{Z}_k$ submanifold of $S^7/\mathbb{Z}_k$. The reduction to Type IIA gives a D4-brane wrapped on a ``squashed'' $CP^2$ inside the $CP^3$ and propagating along the $S^1$ fibre direction\footnote{With an additional D0-brane charge coming from the reduction of the momentum on the orbifold direction, although we have set this to zero in our calculation in section 4.2.}. The ground state corresponds again to the maximal giant, with the squashed $CP^2$ becoming an ordinary $CP^2$ and the direction of propagation collapsing to a point. The configuration is then interpreted as a dibaryon, which arises through our analysis as the limiting case of a D4-brane wrapping a squashed $CP^2$ inside the $CP^3$ with non-vanishing angular momentum.

We should point out that in all the cases that we have considered the branes are effectively wrapped on a $CP^2$ and are therefore subject to the Freed-Witten anomaly \cite{FW}. The branes then carry a half-integer worldvolume magnetic flux that is compensated by a flat half-integer $B_2$-field to produce a vanishing field strength \cite{ABHH}.

We have also addressed the microscopical description in terms of gravitons or D0-branes expanding into fuzzy manifolds, which in all cases involve a fuzzy $CP^2$. This microscopical description is complementary to the macroscopical one in terms of the expanded branes and allows to explore the region with finite 't Hooft coupling. In the microscopical description in terms of 0-branes the worldvolume magnetic flux needed to compensate the Freed-Witten anomaly does not couple in the action, and therefore one has to take into account the effect of the non-vanishing $B_2$. This was studied recently in \cite{LPSS} in the microscopical description of baryon-vertex like configurations in $AdS_4\times CP^3$. There it was shown that the contribution of $B_2$ is subleading in a $1/m$ expansion and has no counterpart in the macroscopical set-up.

We should stress that we have not succeeded in finding a giant graviton expanding on the $CP^3$ in $AdS_4\times CP^3$, other than the well-known dibaryon. The D4-brane wrapping the squashed $CP^2$ inside the $CP^3$ with non-vanishing momentum studied in section 4.2 is described by a Hamiltonian which is very similar to the one describing giant graviton solutions in other backgrounds. However, the ground state corresponds to the maximal giant, i.e. the dibaryon. It may be that inducing angular momentum on the NS5 or D4 brane configurations that we have analyzed by some mechanism similar to that used in \cite{NT} in order to construct the spinning dual giant graviton 
one may be able to obtain the elusive giant graviton solution in $AdS_4\times CP^3$. The fact that our NS5-brane wraps a non-trivial twisted 5-sphere seems to be in agreement with the expectations in \cite{MP}. It would also be interesting to explore the connection with the approach taken in \cite{HMP}, which, based on the similarity between the Klebanov-Witten \cite{KW} and ABJM theories, tries to build a giant graviton solution similar to the D3-brane graviton blown up on the $T^{1,1}$ of the $AdS_5\times T^{1,1}$ background. We hope to report progress in these directions in the near future.

\subsection*{Acknowledgements}

We would like to thank D. Rodr\'{\i}guez-G\'omez  for useful discussions.
Y.L. would like to thank the CERN TH-division, where part of his work was done, for the kind hospitality. This work has been partially supported by the research grants MICINN-FPA2009-07122, MEC-DGI-CSD2007-00042, COF10-03 and the mobility grant MED-PR2010-0476 (Spain).


\begin{thebibliography}{99}

\bibitem{GST}
J.~McGreevy, L.~Susskind, N.~Toumbas, JHEP 0006 (2000) 008, arXiv:hep-th/0003075.

\bibitem{BBNS}
V.~Balasubramanian, M.~Berkooz, A.~Naqvi, M.J.~Strassler, JHEP 0204 (2002) 034, arXiv:hep-th/0107119.

\bibitem{CJR}
S.~Corley, A.~Jevicki, S.~Ramgoolam, Adv. Theor. Math. Phys. 5 (2002) 809, arXiv:hep-th/0111222.

\bibitem{Beren}
D.~Berenstein, JHEP 0407 (2004) 018, arXiv:hep-th/0403110.

\bibitem{MS}
J. Maldacena, A. Strominger, JHEP 9812 (1998) 005, arXiv:hep-th/9804085.

\bibitem{BBFH}
V.~Balasubramanian, D.~Berenstein, B.~Feng, M.X.~Huang, JHEP 0503 (2005) 006, arXiv:hep-th/0411205.

\bibitem{Mello}
R. de Mello Koch, JHEP 0811 (2008) 061, arXiv:0806.0685 [hep-th].

\bibitem{Mello2}
R. de Mello Koch, J.~Smolic, M.~Smolic, JHEP 0706 (2007) 074, arXiv:hep-th/0701066;

R. de Mello Koch, J.~Smolic, M.~Smolic, JHEP 0709 (2007) 049, arXiv:hep-th/0701067;

D.~Bekker, R. de Mello Koch, M.~Stephanou, JHEP 0802 (2008) 029, arXiv:0710.5372 [hep-th].

\bibitem{ABJM}
O. Aharony, O. Bergman, D.L. Jafferis, J. Maldacena, JHEP 0810 (2008) 091, arXiv:0806.1218 [hep-th].

\bibitem{BT}
D.~Berenstein, D.~Trancanelli, Phys. Rev. D78 (2008) 106009, arXiv:0808.2503 [hep-th]

\bibitem{NT}
T.~Nishioka, T.~Takayanagi, JHEP 0810 (2008) 082, arXiv:0808.2691 [hep-th].

\bibitem{BP}
D.~Berenstein, J.~Park, JHEP 1006 (2010) 073, arXiv:0906.3817 [hep-th].

\bibitem{HMPS}
A.~Hamilton, J.~Murugan, A.~Prinsloo, M.~Strydom, JHEP 0409 (2009) 132, arXiv:0901.0009 [hep-th].

\bibitem{HMP}
A.~Hamilton, J.~Murugan, A.~Prinsloo, JHEP 1006 (2010) 017, arXiv:1001.2306 [hep-th].

\bibitem{Dey}
T.K.~Dey, ``Exact Large R-charge Correlators in ABJM theory'', arXiv:1105.0218 [hep-th].

\bibitem{NT2}
T.~Nishioka, T.~Takayanagi, JHEP 0808 (2008) 001, arXiv:0806.3391 [hep-th].

\bibitem{GMT}
M.T.~Grisaru, R.C.~Myers, O.~Tafjord, JHEP 0008 (2000) 040, arXiv:hep-th/0008015.

\bibitem{HHI}
A.~Hashimoto, S.~Hirano, N.~Itzhaki, JHEP 0008 (2000) 051, hep-th/0008016.

\bibitem{SS}
M.M.~Sheikh-Jabbari, J.~Simon, JHEP 0908 (2009) 073, arXiv:0904.4605 [hep-th].

\bibitem{KKL}
J.~Kim, N.~Kim, J.H.~Lee, JHEP 1003 (2010) 122, arXiv:1001.2902 [hep-th].

\bibitem{JLR}
B. Janssen, Y. Lozano, D. Rodr\'{\i}guez-G\'omez, Nucl. Phys. B712 (2005) 371, arXiv:hep-th/0411181.


\bibitem{Myers}
R.C. Myers, JHEP 9912 (1999) 022, arXiv:hep-th/9910053.



\bibitem{PW}
C.N.~Pope, N.P.~Warner, Phys. Lett. B150 (1985) 352.

\bibitem{Pope}
C.N.~Pope, Phys. Lett. B97 (1980) 417.

\bibitem{JL}
B. Janssen, Y. Lozano, Nucl. Phys. B658 (2003) 281, arXiv:hep-th/0207199.


\bibitem{BMN}
D.~Berenstein, J.M.~Maldacena, H.~Nastase, JHEP 04 (2002) 013, arXiv:hep-th/0202021.

\bibitem{LR}
Y. Lozano, D. Rodr\'{\i}guez-G\'omez, JHEP 0508 (2005) 044, arXiv:hep-th/0505073.

\bibitem{JL2}
B. Janssen, Y. Lozano, Nucl. Phys. B643 (2002) 399, arXiv:hep-th/0205254.

\bibitem{JLR2}
B. Janssen, Y. Lozano, D. Rodr\'{\i}guez-G\'omez, JHEP 0611 (2006) 082, arXiv:hep-th/0606264.

\bibitem{JLR3}
B. Janssen, Y. Lozano, D. Rodr\'{\i}guez-G\'omez, JHEP 0706 (2007) 028, arXiv:0704.1438 [hep-th].

\bibitem{TV}
S.P. Trivedi, S. Vaidya, JHEP 0009 (2000) 041, arXiv:hep-th/0007011.

\bibitem{LPSS}
Y. Lozano, M. Picos, K. Sfetsos, K. Siampos, JHEP 1107 (2011) 032, arXiv:1105.0939 [hep-th].

\bibitem{Madore}
J. Madore, ``An Introduction to Noncommutative Differential Geometry and its Applications'', Cambridge University Press, Cambridge, 1995.

\bibitem{dBS}
J. de Boer, K.~Schalm, JHEP 0302 (2003) 041, arXiv:hep-th/0108161.

\bibitem{dBS2}
J. de Boer, K.~Schalm, J.~Wijnhout, Annals Phys. 313 (2004) 425, arXiv:hep-th/0310150.

\bibitem{BFLvR}
D.~Brecher, K.~Furuuchi, H.~Ling, M. van Raamsdonk, JHEP 0406 (2004) 020, arXiv:hep-th/0403289.

\bibitem{BKLvR}
D.~Brecher, P.~Koerber, H.~Ling, M. van Raamsdonk, JHEP 0601 (2006) 151, arXiv:hep-th/0509026.

\bibitem{AJT}
J.~Adam, B.~Janssen, W.~Troost, W. van Herck, Phys. Lett. B662 (2008) 220, arXiv:0712.0918 [hep-th].

\bibitem{BLO}
E.~Bergshoeff, Y.~Lozano, T.~Ort\'{\i}n, Nucl. Phys. B518 (1998) 363, arXiv:hep-th/9712115.

\bibitem{EJL}
E.~Eyras, B.~Janssen, Y.~Lozano, Nucl. Phys. B531 (1998) 275, arXiv:hep-th/9806169.


\bibitem{LR2}
Y. Lozano, D. Rodr\'{\i}guez-G\'omez, JHEP 0608 (2006) 022, arXiv:hep-th/0606057.

\bibitem{FW}
D.S.~Freed, E.~Witten, Asian J. Math. 3 819 (1999) 819852, arXiv:hep-th/9907189.

\bibitem{ABHH}
O.~Aharony, A.~Hashimoto, S.~Hirano, P.~Ouyang, JHEP 1001 (2010) 072, arXiv:0906.2390 [hep-th].


\bibitem{MP}
J.~Murugan, A.~Prinsloo, ``ABJM Dibaryon Spectroscopy'', arXiv:1103.1163 [hep-th].

\bibitem{KW}
I.~R.~Klebanov, E.~Witten, Nucl. Phys. B536 (1998) 199, hep-th/9807080.









\end{thebibliography}
\end{document}